\documentclass[fleqn,usenatbib,letter]{mnras}

\usepackage[T1]{fontenc}
\usepackage{ae,aecompl}
\usepackage{xcolor}

\usepackage{graphicx}	% Including figure files
\usepackage{amsmath}	% Advanced maths commands
\usepackage{amssymb}	% Extra maths symbols

\title[Odd Radio Circles]{On energetics and progenitors of Odd Radio Circles: A causal connection with tidal disruption of stars?}

\author[A. Omar]{
Amitesh Omar\thanks{E-mail: aomar@aries.res.in (AO)}
\\
Aryabhatta Research Institute of observational sciences, Manora Peak, Nainital, 263001, India\\
}

\begin{document}
%\label{firstpage}
%\pagerange{\pageref{firstpage}--\pageref{lastpage}}
\maketitle

% Abstract of the paper
\begin{abstract}
Odd Radio Circles or ORCs are recently discovered edge-brightened, low surface brightness circular radio sources. The progenitors and astrophysical processes responsible for their origins are presently debated. Some ORCs are host-less and some appear to be hosted in distant quiescent galaxies. Two plausible explanations consider ORCs as nearby supernova remnants with sizes a few hundred parsec in the intragroup medium of the local group of galaxies or alternatively shocked halos of a few hundred kpc extent around distant galaxies. The input shock energy required to create ORCs of a few hundred kpc size is estimated in a range of $10^{55} - 10^{59}$ erg. It is shown here that the cumulative energy in unbound debris ejected from multiple ($10^{5} - 10^{9}$) tidal disruption events over $\sim100$ Myr period around a central massive black hole can meet the required energies to generate ORCs around some galaxies, which have recently undergone a merger. The potential hosts for ORCs are identified here as abundant post-starburst galaxies at intermediate redshifts having massive black holes. A causal connection between ORC around quiescent galaxies and tidal disruption may find support in the observed dominance of tidal disruption events in post-starburst galaxies. 

\end{abstract}

% Select between one and six entries from the list of approved keywords.
% Don't make up new ones.

%%%%%%%%%%%%%%%%%%%%%%%%%%%%%%%%%%%%%%%%%%%%%%%%%%

%%%%%%%%%%%%%%%%% BODY OF PAPER %%%%%%%%%%%%%%%%%%

\section{introduction}

Odd Radio Circles (ORCs) are recently discovered low surface brightness (a few hundred $\mu$Jy~beam$^{-1}$) diffuse radio sources at low frequencies ($<1$ GHz), with a nearly circular edge-brightened morphology \citep{Nora, Kori, Fil, Nor22, Oma22}. The ORCs are serendipitous discoveries in the Evolutionary Map of the Universe (EMU; \citealt{Norb}) survey, carried out using the Australian Square Kilometre Array (SKA) Pathfinder (ASKAP) radio interferometric array at $\sim900$ MHz. This survey is carried out at high Galactic latitude regions, where diffuse Galactic radio background emission is weak and the population of Galactic supernova remnants (SNRs) is expected to be very sparse. The ORCs have also been detected from the Giant Meterwave Radio Telescope (GMRT) at 327 MHz \citep{Kori} and Low Frequency Array (LOFAR) radio telescope at 144 MHz \citep{Oma22}. The morphologies of the detected ORCs are strikingly similar to those of the Galactic SNRs. The observed number density of ORCs is about 1 in 50 deg$^{2}$ sky-area. This number density is too large for ORCs to be considered among the known population of the Galactic SNRs, which are found strongly concentrated in the Galactic plane.

In three out of the seven published ORCs, an optical galaxy is found co-located near the geometrical centre of each of the ORCs. The photometric redshifts of the galaxies in these three ORCs are estimated to be 0.55 (ORC-1), 0.46 (ORC-4) and 0.27 (ORC-5) \citep{Nor22}. The colors of these galaxies are on the redder side, hence unlikely to be hosting an ongoing starburst. These three ORCs are likely to be physically associated with the galaxies as the probability of chance alignments between the radio centres of the ORCs and the locations of the galaxies is estimated to be very small. The angular diameters of the ORCs are nearly $1'$. Consequently, the physical extents of these three ORCs are expected in the range of $200-500$ kpc. In the remaining four ORCs, no definite optical counterparts have been identified. The radio emission from some ORCs show a negative spectral index, suggesting that the radio emission is non-thermal and is likely to be  synchrotron radiation.

The astrophysical processes behind ORCs are presently debated with a number of possibilities. The detection of polarization in the radio emission in ORC-1, revealing tangential magnetic field along the ring structure, is interpreted as expanding spherical shock waves \citep{Nor22}. Such a shock wave will be qualitatively similar to a supernova blast wave expanding in the interstellar medium (ISM). Some possibilities tracing origins of the ORCs, as speculated in \citet{Nor22} are - (i) shocks created during mergers of black holes (BH) in galaxies, (ii) shocks created in the winds emanating from star-bursting galaxies (iii) end-on lobes of radio galaxies. The mechanisms behind shocks on a few hundred kpc-scales around normal galaxies are not yet clear, as no definite resemblance with any previously known observed phenomena has been identified so far.

\citet{omar22} presented a possibility that some ORCs can be SNRs in the intragroup medium (IGrM) of the nearest groups of galaxies including the local group. Such SNRs will trace their origins in stars in diffuse tidal streamers in the IGrM, surrounding the Milky Way (MW) and other nearby galaxies. At least one ORC without an optical counterpart has been considered as a SNR in the outskirts of the Large Magellanic Cloud \citep{Fil}. The two other ORCs without an optical counterpart may also be SNRs in the IGrM. Although, only seven ORCs are published, the ORCs as a class of object appear heterogeneous in nature, tracing their origins in distant galaxies or in host-less SNRs in the nearest groups of galaxies. In both the scenarios, the shock wave will be expanding in the IGrM, which is characterized as a low-density medium ($10^{-4}-10^{-5}$ cm$^{-3}$) with high temperature ($10^{6}-10^{7}$ K). The shock expansion energetics in case of ORCs will be different in comparison to that in the SNRs expanding in high-density cold ISM. The differences will arise mainly due to the following reasons - (i) The thermal energy of the swept-up matter in the IGrM becomes significant and as a result the Sedov-Taylor (S-T) phase will terminate earlier in a hotter medium (e.g., \citealt{Tang}). (ii) Due to low density in the IGrM, the final extent of shock towards the end of the S-T phase will be larger. (iii) Due to increase in the sound speed in the hotter medium, the effective Mach number in shocks will be lower, typically in the range of $1-5$, which may reduce  particle acceleration efficiencies in the Fermi process (e.g., \citealt{Hoeft}). 

It is worth to note that nearly-spherical shocks are possible in number of astrophysical scenarios other than the SNRs. \citet{Gil} described shocks caused by the outflows around growing BH by continuous or intermittent accretion in a context of high redshift quasars. The shocks are also predicted in tidal disruption event (TDE) as a star enters within the tidal radius of a massive BH and is disintegrated \citep{Alex}. The shock energies in a TDE event can be comparable to that in a supernova explosion \citep{Kho}. In this paper, shock energetics of multiple TDE events resulting from continuous capture of stars by a massive BH in a galaxy is compared with the required input shock energies for the creations of ORCs. The motivation for studying TDEs in the context of ORCs comes from the observed dominance of TDEs in post-starburst (PSB) galaxies (see e.g., \citealt{Bor}) and some remarkable similarities noticed between the properties of the galaxies associated with some ORCs and those of the PSBs. 

\section{Energetics of ORC} 

It is assumed here that the ORCs are expanding shock waves in low-density hot environments of IGrM. The particle acceleration to the ultra-relativistic energies takes place in shocks via the first-order Fermi process and the TeV electrons emit the radio synchrotron emission at GHz frequencies in presence of a few $\mu$G magnetic fields. The host-less ORCs, which can be identified as SNRs in the IGrM are expected to have their sizes a few hundred parsec while the ORCs associated with distant galaxies are expected to have their sizes of the order of a few hundred kilo-parsec. The general framework of expanding shocks, extensively discussed in the published literature in contexts of SNRs will be followed here. As the solutions in the S-T phase for the expanding shocks are self-similar, it can be scaled. 

\subsection{ORC as IGrM supernova remnant} 

\citet{omar22} showed that a measurable fraction of ORCs may be identified as SNRs in the IGrM, associated with the local group and its immediate neighbour groups of galaxies. The arguments and estimates in \citet{omar22} are based on the optical detection rate of the host-less (not within a galaxy) supernova in a large number (1401) of nearby groups of galaxies, constrained using the Sloan Digital Sky Survey (SDSS) supernova survey database by \citet{Mcgee}. It was found that angular size, surface brightness and radio flux of the host-less SNRs within a distance of $\sim3$ Mpc in the IGrM can be similar to the known ORCs. It was estimated that nearly 5400 IGrM supernova are expected to take place in the local group and its immediate neighbour groups of galaxies in $\sim1$ Myr period. 

The radio detectability periods of the remnants of IGrM supernova can be estimated based on some assumptions. If one assumes that the majority of SNRs are detected around the end of their S-T phase, the Eq. 3 of \citet{Tang} can be used to estimate this characteristic time. For an IGM temperature of $10^{7}$ K, this time is $\sim0.24$ Myr and $\sim0.52$ Myr for a SNR expanding in an ambient density of $10^{-4}$ cm$^{-3}$ and $10^{-5}$ cm$^{-3}$ respectively. For a lower IGrM temperature ($10^{6}$ K), this time is estimated to be between $1.5-3.5$ Myr for the same range of the ambient density as before. Due to a lower ambient density in the IGrM compared to that in the ISM, the SNR can grow up to a few hundred parsec in size in typical IGrM environments. Although, the predicted S-T phase is sufficiently long, the radio surface brightness of SNRs may fall below the detection limits ($\sim10^{-22}$ W~m$^{-2}$~Hz$^{-1}$~sr$^{-1}$) of the present radio surveys near 1 GHz. Therefore, while a IGrM-SNR may dynamically exist for a Myr period, the radio detectability will require higher surface brightness sensitivities. As only a few Galactic SNRs are known in low-density warm medium in the ISM, it is difficult to predict detectability of SNRs in a warm-hot IGrM. \citet{omar22} assumed a radio detectability of $10^4$ years to estimate observable SNR number density as nearly 54 SNRs or 1.3 SNR per 1000 square degree in the IGrM. This estimate can easily vary by a factor of a few depending upon the number of supernova progenitors at an epoch and the radio detectability periods of SNRs in different IGrM environments. 

The purpose of the estimate made here was to check whether the predicted numbers of IGrM-SNRs under some realistic assumptions are significant and whether SNRs can represent a measurable population of the ORCs. It appears that at least those ORCs, which are detected without optical counterparts, may be SNRs in the IGrM. 

\subsection{ORC as shocks from TDEs in a galaxy}

If ORCs are shocked regions on a few hundred kpc-scales around galaxies, the energy required to create such shocks can be estimated following the relation given in \citet{Blon}. The minimum required energy ($E$) to produce ORC of a radius $r$ (pc) towards the end of the S-T phase in an ambient medium density $n$ (cm$^{-3}$) can be estimated using the relation: 

\begin{equation}
E (10^{51}~erg) \approx (0.05r)^{17/5}~n^{7/5} 
\end{equation}

The estimated energy is nearly $10^{57}$ erg and $2\times10^{59}$ erg for the ORCs of sizes 100 kpc and 500 kpc respectively, expanding in an ambient density of $10^{-4}$ cm$^{-3}$. In lower ambient density of $10^{-5}$ cm$^{-3}$, the required energy will be nearly $4\times10^{55}$ erg and $10^{58}$ erg for ORCs of sizes 100 kpc and 500 kpc respectively. We also estimated the energetics using the S-T solutions provided in \citet{Aved}, for an analogues astrophysical context of continuous energy deposition by the stellar winds from massive stars in the ISM. The estimates from this approach and that using Eq.~1 agree within a factor of a few. The expected Mach numbers are between 1 and 2 towards the end of the S-T phase. Typical ages (characteristic time at the end of the S-T phase) of such ORCs are estimated to be $10 - 1000$ Myr for the range of density and temperature of the IGrM explored in this paper. Again, the detectability of ORCs within the sensitivities of the present surveys may limit detections of ORCs in a younger phase.

The energy released in TDEs can be estimated and compared with the required energy budgets of ORCs to check if TDEs can provide required power for ORCs. When a star enters within the tidal radius of a massive BH, it gets disintegrated and a TDE may take place \citep{Rees}. The TDEs can take place around a BH with mass $<10^{8}$ M$_{\odot}$. The energetics and initial conditions for TDEs are described in the review article by \citet{Alex}. The energy in this process is released via two paths - (i) a luminous accretion flare lasting a few years by the captured mass surrounding the massive BH, (ii) a shock-wave analogous to the supernova blast wave from nearly half of the disrupted star's mass as unbound ejecta. Using the relation given in \citet{Kho}, the energy deposited in the ISM via shock waves from the unbound ejecta can be as high as $\sim10^{53}$ erg for a solar mass star tidally disrupted by the BH with a mass $4\times10^{6}$ M$_{\odot}$ residing in the Galactic Centre (GC). This estimate assumed deeply penetrating orbits. However, \citet{Gui} showed by simulations that typical energy deposited  will be significantly lower and will have a median value of nearly $5\times10^{49}$ erg in case of tidal disruptions of solar mass star by the GC-BH. For a massive main sequence star with mass 30 M$_{\odot}$, the energy is estimated to be $\sim30$ times higher considering a mass-radius relation for stars given in \citet{Eker} and tidal disruption theory outlined in \citet{Rees} and \citet{Gui}. This energy scales with the mass of the BH as $M_{BH}^{1/3}$. The important consequence of this result can be that the energy has only a weak dependence on the mass of the BH and hence TDEs around relatively less massive BH (e.g., $10^{4}-10^{5}$ M$_{\odot}$) can also contribute significantly to the total energy deposition by the unbound ejecta. For instance, a TDE around a BH mass of $10^{5}$ M$_{\odot}$ will provide energy only about a factor of 3.5 lower than that by the GC-BH. Therefore, majority of galaxies with massive ($10^{4} - 10^{8}$ M$_{\odot}$) BH can be potential candidates for TDEs. The energy deposited by the unbound ejecta can not be constrained as these are the long-term effects of tidal disruptions, which have not been observationally verified in individual TDEs. In this paper, a fiducial average value of $\sim10^{50}$ erg of released energy per TDE is used.

In order to meet the required energy ($10^{55} - 10^{59}$ erg) budget to power ORCs of a few hundred kpc size, multiple TDEs of the order of $10^{5} - 10^{9}$ are needed in a time shorter than the maximum age ($\sim100$ Myr) of  the relativistic electrons in ORCs. Although, typical TDE rates in galaxies similar to the MW are expected to be of the order of $10^{-4} - 10^{-5}$ per year \citep{Kho, Wang}, the rates can be dramatically enhanced up to a value of $\sim1$ event per year after a merger event between two galaxies (e.g., \citealt{Cen} and \citealt{Madigan}). \citet{Cen} argued that a secondary massive BH inspiraling through the galaxy's disk as a result of a merger of two galaxies will be responsible for the enhanced TDE rate. \citet{Madigan} showed via simulation of stellar orbits in eccentric nuclear disks created as a result of merger that such eccentric disks can remain stable and result into large number of stars crossing tidal sphere of a massive BH. Both the above mechanisms may explain the preferential occurrence of TDEs in the post-merger PSB galaxies (e.g. see \citealt{Bor} and references therein).

The PSB galaxies are presently not forming new stars but these galaxies have very likely undergone a starburst within the past $0.1-1$ Gyr period as indicated by their optical spectra. A trigger of intense starburst is normally traced to merger of galaxies both in massive systems (e.g., \citealt{Mihos}) as well as in dwarf galaxies (e.g., \citealt{sumit}). Therefore, the PSB galaxies can be natural hosts of TDEs. Considering all possible favourable conditions for a enhanced rate ($\sim$1 per year) of TDEs in the PSB galaxies, the total energy deposition rate via shock-wave from the unbound ejecta from TDEs can reach $10^{58}$ erg over $\sim100$ Myr period. This rate can be expected to decline over time, however this decline is predicted to be slow and a high TDE rate may be sustained for up to 1 Gyr after the starburst in certain conditions (see e.g., \citealt{Bor}). The important outcome of this analysis is that there exists possibilities where a enhanced rate of TDEs in PSB galaxies can make large energy deposition, matching with those required to generate some ORCs on hundred-kpc scales. 

It is also worth to examine if the energy deposition rate from multiple TDEs can be sufficient for shocks to penetrate the dense nuclear regions and breakout in the IGrM to create shocked radio halos. \citet{Ko} estimated that a minimum rate of $\sim10^{41}$ erg s$^{-1}$ is required for shocks to penetrate through the central dense molecular disk (density $<n>\sim50$ cm$^{-3}$) of thickness a few tens of parsec in the GC in order to drive shocks into the Galactic halo. This minimum rate of energy deposition can be easily met by the enhanced rate (1 per year) of TDEs ($\sim10^{50}$ erg per event) generating nearly $3\times10^{42}$ erg s$^{-1}$ energy. Therefore, this mechanism is likely to contribute even if a TDE produces an order of magnitude less energy than the assumed value of $\sim10^{50}$ erg or the TDE rate is low by an order of magnitude. 

The effect of injecting energy to ISM and IGrM via multiple shocks from the unbound ejecta of TDEs in a short span of time can be expected to be energetically similar to that due to multiple supernova explosions in the nuclear regions of the starburst galaxies, where superwinds and superbubbles can be created (e.g., see \citealt{Sharma}). In these superwinds, particles can be accelerated to the relativistic energies \citep{Romero}. These relativistic particles will emit synchrotron radiation in radio bands in presence of magnetic fields. 
 
\section{PSB galaxies as potential hosts of ORC}

Since detections of TDEs have been observed to be dominating in the PSB galaxies, it is worth to discuss population density and environments of the PSB galaxies. The spectra of PSB galaxies show strong Balmer absorption lines, indicators of presence of recently formed stars, with ages within $\sim1$ Gyr (see e.g., \citealt{French}). At the same time, the PSB galaxies do not have strong emission lines indicating lack of ongoing star formation. The number density of the PSB galaxies is known to evolve strongly with the redshift between 0 and 1. The analyses of a large number of PSB galaxies selected from the SDSS and Baryon Oscillation Spectroscopic Survey (BOSS) showed that the number density of the PSB galaxies at fixed absolute magnitude, e.g. at -23 mag (at 500 nm rest-frame; at $z\sim0.5$ near the i-band), varies by more than an order of magnitude between the redshift 0.2 and 0.6, with higher density at $z\sim0.6$ \citep{Patta}. The space density of the PSB galaxies ($M_*>10^{9.75}$ M$_{\odot}$) at $z\sim0.7$ is about $10^{-4}$ Mpc$^{-3}$, while that at $z\sim0.07$ is $\sim5\times10^{-7}$ Mpc$^{-3}$ \citep{Wild}. 

If ORCs are primarily hosted by the PSB galaxies, the chance findings of the ORCs can be expected to increase significantly at intermediate redshifts, due to increased number density of the PSB galaxies in those epochs. Although, it can be too early to draw any significant conclusion with just three ORCs with approximate photometric redshifts, a strong evolution of the PSB galaxies with the redshift may be a reason why three ORCs with host galaxies are found at intermediate redshifts in the range $z\sim0.27-0.55$. The presently constrained space density of the ORCs is about $2\times10^{-8}$ Mpc$^{-3}$ \citep{Nor22}, far below the space density of the PSB galaxies. If a connection between ORCs and PSB galaxies is real, the presently detected population of ORCs under the sensitivity limit of the ASKAP survey may therefore only be a tip of the iceberg with many more ORCs waiting to be detected in deep surveys with order of magnitude higher sensitivity with the full SKA. A word of caution is also necessary here, as one should not expect all PSB galaxies to host ORCs, since it requires several favourable conditions to produce large number of energetic TDEs in a galaxy. We presently do not know how often all favourable conditions to generate ORCs can be met in a merger system of galaxies. 

As only one ORC namely ORC-1 has been extensively studied, we can highlight some properties of the host galaxy of the ORC-1 in the context of a possible connection between the PSB galaxies and ORCs proposed in this paper. The host galaxy of the ORC-1 ($z\sim0.55$; $M_*\sim2.7\times10^{11}$ M$_{\odot}$; $M_i\sim-23$ mag) reside in a over-dense environment. The other two ORCs at $z\sim0.27$ and $z\sim0.46$ although do not reside in a over-dense environment but have a few neighbour galaxies. This observation led \citet{Nor22} to consider a possible role of galaxy interactions in ORCs. A study of environments of the PSB galaxies using the SDSS and DEEP2 redshift surveys reveals that the PSB galaxies at $z\sim0.8$ reside in over-dense environments while those at $z\sim0.1$ prefer under-dense environments \citep{Yan}. This indicates that both number density and environment of the PSB galaxies vary strongly with the redshift between 0 and 1. A spectral energy distribution fit to the ORC-1 host galaxy also suggests that it is presently not forming stars but shows evidences of an intense star-burst several Gyr ago \citep{Nor22}. This observation is similar to that in the PSB galaxies. The ORC-1 also has an active galactic nuclei  with 1.4 GHz radio luminosity of $7.2\times10^{22}$ W Hz$^{-1}$, which can be a strong indicator of the presence of a massive BH in the galaxy. Several properties of the host galaxy of the ORC-1 therefore broadly match with the scenario presented here in which ORCs are predicted to be hosted in PSB galaxies with a massive BH.   

\section{Summary and concluding remarks}

In this paper, PSB galaxies are proposed to be the potential hosts for some ORCs, recently detected in low frequency deep radio surveys. The expanding shocks from the unbound ejecta of a large number of TDEs taking place in some PSB galaxies is proposed to be generating radio halos. This proposition finds some support from the known properties of the PSB galaxies. A summary of the paper is presented below.  

\begin{itemize}
\item
Two possibilities were discussed here - (i) Some ORCs as SNRs in the IGrM associated with the local group and its immediate neighbour groups of galaxies as proposed by \citet{omar22}. (ii) Some ORCs hosted by PSB galaxies with a massive BH, where TDEs are occurring at an enhanced rate due to a recent merger between galaxies.  
\item
The energetics of the shocks ($\sim10^{58}$ erg) resulting from a high rate (1 per year) of TDEs over a 100~Myr period was shown here to broadly agree with the  energy requirements ($10^{55} - 10^{59}$ erg) for the ORCs, estimated using the S-T solutions of expanding shocks in a low-density hot environment typical of IGrM. This estimate has ample margin so that some ORCs around quiescent galaxies can still be explained if the TDE rate or the average energy output per TDE is order of magnitude smaller.
\item
Two observed properties of the PSB galaxies, namely (i) high space density at intermediate redshifts and (ii) prefer to reside in over-dense environments at intermediate redshifts together with the higher incidences of the observed TDEs in the PSB galaxies, support the proposed scenario for some ORCs being hosted in the PSB galaxies. 
\item
The observed properties of the well-studied ORC-1 host galaxy at $z\sim0.55$, - (i) residing in an over-dense environment, (ii) spectral energy distribution analysis suggestive of a starburst a few Gyr ago, (iii) presence of a radio AGN hence that of a massive BH in the galaxy, agree well with the expected properties for the host galaxies of ORCs.
\item
The chance findings of ORCs at intermediate redshifts ($z\sim0.27$ to 0.55) agree well with the expected trend from the higher observed densities of the PSB galaxies at intermediate redshifts. 
\item
If a causal connection between PSB galaxies, TDEs and ORCs gets firmly established, it is likely that a large number of ORCs will be detected in future with high sensitivity survey using the LOFAR and SKA telescopes.
\end{itemize}

We also make some remarks here for extending this research work in future -\\ 

(1) The GC has shocked bubbles on scales of a few kpc (termed as Fermi bubbles) emanating outwardly from the two opposite sides of the Galactic plane. The Fermi bubbles have been detected in the Gamma-rays \citep{DobFB} and also in the radio bands \citep{Dob}. Multiple periodic TDEs in the GC were proposed to be behind the creation of the Fermi bubbles \citep{Cheng, Ko}. The ORCs may be the extra-galactic analogues of the Galactic Fermi bubbles but on larger scales with higher energies. 

(2) The disrupted mass in a TDE is ejected mostly on one half of the orbital plane, in contrast with a spherical supernova explosion \citep{Kho}. Hence, multiple shocked geometries in different orientations can be expected from multiple TDEs. The high angular resolution radio image of the ORC-1 as presented in \citet{Nor22} revealed two prominent rings with some significant arc-like structures, which could have resulted from multiple bursts of TDEs. It may be noted that the diffusion of the TeV electrons responsible for the radio emission can be slow ($\sim0.1$ kpc Myr$^{-1}$; see e.g., \citealt{omar19}) in the IGrM environments, hence, the individual shock fronts at the end of the S-T phase can remain visible for a few tens of Myr period.

(3) The ORCs have potential to become important tracers of activities related to the massive BH in galaxies, particularly in the context of a preposition first proposed by \citet{Gopal}, where TDEs in galaxies with BH mass $<10^{8}$ M$_{\odot}$ may explain the well-known radio galaxy dichotomy based on radio loudness. 

\section*{Acknowledgements}
Author acknowledges stimulating discussions with Prashant Mohan, Jagdish Joshi, Ravi Joshi and Koshy George. We thank the referee for constructive comments. 

\section*{data availability} 
No new data were generated in support of this research. All the data underlying this article are available in the article or from the peer-reviewed publications enumerated in the reference section.

% Don't change these lines
\bsp	% typesetting comment
\label{lastpage}

\begin{thebibliography}{99}

\bibitem[\protect\citeauthoryear{Alexander}{2005}]{Alex}
Alexander T., 2005, PhR, 419, 65
\bibitem[\protect\citeauthoryear{Avedisova}{1972}]{Aved}
Avedisova V. S., 1972, SvA, 15, 708
\bibitem[\protect\citeauthoryear{Blondin et al.}{1998}]{Blon}
Blondin J. M., Wright E. B., Borkowski K. J., Reynolds S. P. 1998, ApJ, 500, 342
\bibitem[\protect\citeauthoryear{Bortolas}{2022}]{Bor}
Bortolas E., 2022, MNRAS, 511, 2885
\bibitem[\protect\citeauthoryear{Cen}{2020}]{Cen}
Cen R., 2020, ApJL, 888, L14
\bibitem[\protect\citeauthoryear{Cheng et al.}{2011}]{Cheng}
Cheng K.-S., Chernyshov D. O., Dogiel V. A., Ko C.-M., Ip W.-H., 2011, ApJL, 731, L17 
\bibitem[\protect\citeauthoryear{Dobler}{2010}]{DobFB}
Dobler G. et al., 2010, ApJ, 717, 825
\bibitem[\protect\citeauthoryear{Dobler}{2012}]{Dob}
Dobler G., 2012, ApJ, 750, 17
\bibitem[\protect\citeauthoryear{Eker et al.}{2018}]{Eker}
Eker Z. et al., 2018, MNRAS, 479, 5491
\bibitem[\protect\citeauthoryear{Filipovic et al.}{2022}]{Fil}
Filipovic M. D. et al., 2022, MNRAS, 512, 265
\bibitem[\protect\citeauthoryear{French}{2021}]{French}
French K. D., 2021, PASP, 133, 072001
\bibitem[\protect\citeauthoryear{Gilli et al.}{2017}]{Gil}
Gilli R., Calura F., D’Ercole A., Norman C., 2017, A\&A, 603, A69 
\bibitem[\protect\citeauthoryear{Gopal-Krishna, Mangalam \& Wiita}{2008}]{Gopal}
Gopal-Krishna, Mangalam A., Wiita P. J., ApJL, 2008, 680, L13
\bibitem[\protect\citeauthoryear{Guillochon et al.}{2016}]{Gui}
Guillochon J., McCourt M., Chen X., Johnson M. D., Berger E., 2016, ApJ, 822, 48
\bibitem[\protect\citeauthoryear{Hoeft \& Bru\.g\.gen}{2007}]{Hoeft}
Hoeft M., Bru\.g\.gen M., 2007, MNRAS, 375, 77
\bibitem[\protect\citeauthoryear{Jaiswal \& Omar}{2020}]{sumit}
Jaiswal S., Omar A., MNRAS, 2020, 498, 4745
\bibitem[\protect\citeauthoryear{Mihos \& Hernquist}{1996}]{Mihos}
Mihos J. C., Hernquist L., ApJ, 464, 641
\bibitem[\protect\citeauthoryear{Khokhlov \& Melia}{1996}]{Kho}
Khokhlov A., Melia F., 1996 ApJL, 457, L61
\bibitem[\protect\citeauthoryear{Ko et al.}{2020}]{Ko}
Ko C. M., Breitschwerdt D., Chernyshov D. O., Cheng H., Dai L., Dogiel V. A., 2020, ApJ, 904, 46
\bibitem[\protect\citeauthoryear{Koribalski et al.}{2021}]{Kori}
Koribalski B. S., Norris R. P., Andernach H., Rudnick L.,  Shabala S., Filipovic M., Lenc E., 2021, MNRAS, 505, L11
\bibitem[\protect\citeauthoryear{Madigan et al.}{2018}]{Madigan}
Madigan A.-M., Halle A., Moody M., McCourt M., Nixon C., Wernke H., 2018, ApJ, 853, 141
\bibitem[\protect\citeauthoryear{McGee \& Balogh}{2010}]{Mcgee} 
McGee S. L., Balogh M. L., 2010, MNRAS, 403, 79
\bibitem[\protect\citeauthoryear{Norris et al.}{2021a}]{Nora}
Norris R. P. et al., 2021a, PASA, 38, 3
\bibitem[\protect\citeauthoryear{Norris et al.}{2021b}]{Norb}
Norris R. P. et al., 2021b, PASA, 38, e046
\bibitem[\protect\citeauthoryear{Norris, Crawford \& Macgregor}{2021}]{Norc} 
Norris R. P., Crawford E., Macgregor P., 2021, Galaxies, 9, 83
\bibitem[\protect\citeauthoryear{Norris et al.}{2022}]{Nor22}
Norris R. P. et al., 2022, MNRAS, 513, 1300
\bibitem[\protect\citeauthoryear{Omar}{2019}]{omar19}
Omar A., 2019, MNRAS, 484, L141
\bibitem[\protect\citeauthoryear{Omar}{2022}]{omar22}
Omar A., 2022, MNRAS, 513, L101
\bibitem[\protect\citeauthoryear{Omar}{2022}]{Oma22}
Omar A., 2022, Res. Notes AAS, 6, 100
\bibitem[\protect\citeauthoryear{Pattarakijwanich et al.}{2016}]{Patta}
Pattarakijwanich P. et al., 2016, ApJ, 833, 19
\bibitem[\protect\citeauthoryear{Pavlovi\'c et al.}{2018}]{pavl}
Pavlovi\'c, M. Z., Uro\'sevic, D., Arbutina B., Orlando S., Maxted N., Filipovic M. D., 2018, ApJ, 852, 84
\bibitem[\protect\citeauthoryear{Rees}{1988}]{Rees}
Rees M. J., 1988, Nature, 333, 523
\bibitem[\protect\citeauthoryear{Romero, M\"uller \& Roth}{2018}]{Romero}
Romero G. E., M\"uller A. L., Roth M., 2018, A\&A, 616, A57
\bibitem[\protect\citeauthoryear{Sharma et al.}{2014}]{Sharma}
Sharma P., Roy A., Nath B. B., Shchekinov Y., 2014, MNRAS, 443, 3463
\bibitem[\protect\citeauthoryear{Tang \& Wang}{2005}]{Tang} 
Tang S., Wang Q. D., 2005, ApJ, 628, 205 
\bibitem[\protect\citeauthoryear{Wang \& Merritt}{2004}]{Wang}
Wang J., Merritt D., 2004, ApJ, 600, 149
\bibitem[\protect\citeauthoryear{Wild et al.}{2009}]{Wild}
Wild V. et al., 2009, MNRAS, 395, 144
\bibitem[\protect\citeauthoryear{Yan et al.}{2009}]{Yan} 
Yan R. et al., 2009, MNRAS, 398, 735
\end{thebibliography}
\end{document}